\documentclass{article}
\usepackage[english]{babel}
\usepackage[a4paper]{geometry}

\parskip=\smallskipamount
\advance\textwidth by -2cm
\advance\oddsidemargin by 1cm
\advance\evensidemargin by 1cm

\title{Stanford Matrix Considered Harmful}
\author{Sebastiano Vigna}
\date{}
\begin{document}

\maketitle

The idea for this note arose\footnote{Actually, 
I must thank Michela Redivo Zaglia, who challenged me to write it.} during the
``Web Information Retrieval
and Linear Algebra Algorithms'' held at Schloss Dagstuhl in February 2007. Many
brilliant people working on either side (numerical analysis and web search) had a chance
to meet and talk for one week about mathematical and practical aspects of linear
methods for ranking, and in particular (not surprisingly) PageRank and
HITS.\footnote{This note contains intentionally no references.
After some reckoning, this was the only sensible 
choice, as this note is intended to promote critical 
discussion instead of providing information for 
citation-based importance measures. No claim in this note is new---the facts I
report are easily found in the literature.}

There were many scientific aspects that were surprising for both sides and that
were (finally, one might say) stated clearly at the workshop. First of all,
PageRank is \emph{not} the most important factor in Google's scoring, or in any
search engine scoring. It is part of literally hundreds of features that are
somehow combined (e.g., by a standard machine-learning framework), and its
importance has decreased in time 
(people in the so-called ``search engine optimisation" industry claim that
the importance of PageRank dropped drastically around 2003, but this 
claim is
based just on reverse engineering). Yet papers are published every day
starting with claims such as ``PageRank is the most important ranking\dots''
and so on. Such statements give a distorted view of reality, but clearly work
very well for publishing papers.

On the other hand, it is also clear that PageRank (and, more generally,
link-based ranking) is useful in other areas, such as deciding which pages to
crawl and which not. 

Moreover, when one attacks the \emph{real} PageRank computation problem 
suddenly many sophisticated methods devised by the numerical analysis community 
are pretty useless because of the very large size of the matrix. This problem 
actually had the very positive effect of stimulating a lot of new  research in 
numerical analysis methods that could be applied to large, sparse, irregular 
stochastic matrices of dimension $10^{10}$ and beyond.  Much of the classical 
work on such large systems involved matrices with considerable structure from 
ODEs, PDEs, queuing theory, etc. Actually, many of the more promising 
approaches for PageRank computation use some kind of graph structural analysis 
(usually related to strongly connected components) that is rather meaningless 
in the classical problems mentioned above.

These considerations bring us to the point of this note. The problem of computing
PageRank is interesting from a practical viewpoint only if the \emph{size} of the matrix
is large and if the \emph{type} of the matrix is a web graph. 

What do we mean by ``large''?
Currently, search engines claim to index a number of pages in the order of
$10^{10}$. We cannot expect, as scientists, to replicate exactly what an infrastructure
with a billion-dollars budget does, but let's say that we can be indulgent with
ourselves and reduce our problem by three orders of magnitude.

Defining the type of a web-graph matrix is a much more subtle issue. However, if
claiming that a matrix is \emph{like} a real web graph might be difficult,
claiming that is \emph{definitely not like} a real web graph is much
easier---just look for a significant divergence in basic statistical facts known
for web graphs.

The so-called Stanford web matrix is a very small crawl of the Stanford (and
maybe surrounding) sites freely available on the web. The matrix is easily
available in a sparse format that is immediately usable on off-the-shelf
commercial computational tools, so it has become somehow a benchmark, and often
\emph{the} matrix of choice for testing algorithms or statistical hypotheses
about the web. A considerable set of papers in the
literature use this
matrix to perform experiments and to derive conclusions from.

That matrix has less then 300\,000 nodes, and it is thus
\emph{five orders of magnitude} smaller than real-world web snapshots. There are
freely available tools and datasets that make it possible to compute PageRank on
a laptop for snapshots with 100 million nodes in few hours: there is no excuse for
using such a poor sample.

But size is not the only problem: when we look at absolutely elementary
statistics, the Stanford matrix is a complete outlier. As a comparison, I have
used a small (one million nodes) crawl of the \texttt{.eu} domain, a slightly
larger crawl ($7.4$ million nodes) of Indochina countries, and, finally, two 
significantly larger crawls of 40 million and 100 million nodes of the
\texttt{.uk} domain (all these snapshots are publicly available).

As a first comparison, let's look at the density. The most sensible
measure we can use is the average outdegree (proper density is ludicrously small
in web graphs).
It is in the range $22$-$32$ for all involved graphs, except for the Stanford
Matrix, which has average outdegree $8.2$.

A fundamental issue in web crawling is that nobody can crawl the whole web, so
the successors of a significant fraction of pages are not visited. This
causes a significant percentage of the nodes to be \emph{dangling} (i.e., to
have no outlinks). Patching dangling nodes to make the matrix stochastic is a
subtle
issue, and there are unfortunately at least two distinct versions of PageRank
around with different solutions. One can even forget about the Markov chain and
just compute PageRank by solving the associated linear system as if there were
not dangling nodes: the result changes only by a scale factor, but the factor
depends on how many dangling nodes there are, so threshold on the norm of the
vector used to stop approximations has a completely different meaning in this
case. 

The percentage of dangling nodes is in the range $8$-$17$ in all our
examples, except for the Stanford matrix, where it is $0.06$ [sic]. The
resulting linear system will have a radically different behaviour than any real
snapshot.

Another relevant class of nodes is that of \emph{buckets}. A \emph{bucket
component} is a component that is terminal, but not dangling, in the component
DAG. A \emph{bucket (node)} is a node whose component is a bucket component.
When the damping factor $\alpha$ appearing in the definition of PageRank goes to
one, all
rank concentrates in the buckets, and the remaining nodes go to rank zero. This
phenomenon is very important, as it implies that when $\alpha\to1$ PageRank is
essentially meaningless (e.g., the entire core component will get rank zero):
variants of PageRank suited for $\alpha\to1$ should be computed instead (but you
can actually find many papers trying to compute PageRank in the $\alpha\approx1$
region). The graphs under examination have about $3$-$12$\% of buckets, except for the
Stanford matrix, with a whopping 41\% buckets.

All these measures confirm that statistically the Stanford matrix does not look
at all as a web graph. But let's go further: real web graphs have a kind of
\emph{self-similarity} that is exploited by all approaches to web-graph
compression. In particular, several nodes have similar successor lists, and a
simple way to single out such nodes is to permute the graph so to order
lexicographically nodes by their URL. We cannot perform such an experiment
because URLs are not available for the Stanford matrix. But we can do better:
we can permute the nodes by their Gray codes---each row is interpreted
as a tuple of zeroes and ones, and the nodes are reordered following a standard
recursive Gray code. The resulting matrices exhibit similarity in a
coordinate-free way (i.e., without referring to URLs).

When we compress a graph using Gray codes and the WebGraph framework standard
parameters, we obtain about $2$-$3$ bits per link in all the samples
discussed, except for the \texttt{.eu} graph, which requires $4.9$ bits per
link. The Stanford matrix requires $13.8$ bits per link.

There is an interesting phenomenon going on: some typical properties (e.g., high
compressibility) arise in our examples only beyond a certain size (about 10
million nodes). People invoking the ``fractal nature'' of the web as an excuse
to use small samples should thus be very careful (the \texttt{.eu} snapshot, for
instance, is not a very good candidate).

Note that I am not suggesting that all web graphs should look the same, or that
we should set up some standards to define a web graph: there is a healthy
diversity of structure in the real world due to culture, wealth, and available
tools (content-management systems, for instance, have steadily increased the
average outdegree of the web in the last 5 years). But there are criteria, based on
common sense and experience, that should delimit what we use in our experiments
if we want to derive sensible conclusions, and the Stanford matrix largely falls
short of such criteria.

\textit{Post scriptum.} David Gleich (whose most useful comments I'm happy 
to acknowledge) made me notice that the computer science
community is not providing data and tools in a language and in a format that the 
numerical analysis community can easily exploit,
which makes using many realistic, available data sets very difficult.

We're working on that.

\end{document}